\documentclass[10pt, twocolumn]{IEEEtran}
\usepackage{optidef}
\usepackage{longtable}
\usepackage{color}
\usepackage{amsmath}
\usepackage{cancel}
\usepackage{amssymb}
\usepackage{graphicx}
\usepackage{longtable}
\usepackage{multirow}
\usepackage{stfloats}
\usepackage{array}
\usepackage[labelsep=period]{caption}
\usepackage{setspace}
\usepackage{float}
\usepackage{subfig}
\usepackage{lettrine}
\usepackage[noadjust]{cite}
\usepackage{balance}
\usepackage[english]{babel}
\usepackage[letterpaper, left=0.625in, right=0.625in, bottom=1in, top=0.7in]{geometry}
\usepackage[figurename=Fig.]{caption}
\usepackage{suffix}
\usepackage{mathtools}
\usepackage[dvipsnames]{xcolor}

\DeclarePairedDelimiterX\MeijerM[3]{\lparen}{\rparen}%
{#3\delimsize\vert\,\begin{smallmatrix}#1 \\ #2\end{smallmatrix}}

\newcommand\MeijerG[8][]{%
	G^{#2,#3}_{#4,#5}\MeijerM[#1]{#6}{#7}{#8}}

\WithSuffix\newcommand\MeijerG*[7]{%
	G^{#1,#2}_{#3,#4}\MeijerM*{#5}{#6}{#7}}
\def\BibTeX{{\text B\kern-.05em{\sc i\kern-.025em b}\kern-.08em
		T\kern-.1667em\lower.7ex\hbox{E}\kern-.125emX}}
\usepackage{cleveref}
\begin{document}
	\title{Outage Performance of Fluid Antenna System (FAS)-aided Terahertz Communication Networks}
	\author{Leila Tlebaldiyeva, 
		$^{\circ}$Sultangali Arzykulov, 
		$^{\#}$Khaled M. Rabie, 
		$^{\ddagger}$Xingwang Li, and 
		Galymzhan Nauryzbayev
		\\% <-this % stops a space
		%\vspace{0.25cm}
		\IEEEauthorblockA{\normalsize School of Engineering and Digital Sciences, Nazarbayev University, Astana, Z05H0K3, Kazakhstan\\
			$^{\circ}$CEMSE Division, King Abdullah University of Science and Technology, Thuwal, KSA 23955-6900\\
			$^{\#}$Department of Engineering, Manchester Metropolitan University, Manchester, M15 6BH, UK\\ 
			$^{\ddagger}$School of Physics and Electronic Information Engineering, Henan Polytechnic University, Jiaozuo 454000, China \\
			Emails: \{ltlebaldiyeva, galymzhan.nauryzbayev\}@nu.edu.kz, $^{\circ}$sultangali.arzykulov@kaust.edu.sa, \\ $^{\#}$k.rabie@mmu.ac.uk, 
			$^{\ddagger}$lixingwang@hpu.edu.cn
		}
	}
	\maketitle
	\thispagestyle{empty}
\begin{abstract}
\boldmath
Millimeter-wave networks have already been successfully rolled out in many countries and now the research direction heads toward new technologies and standards to enable Tbps rates for future sixth-generation (6G) wireless communication systems. This work studies a point-to-point terahertz (THz) communication network exploiting the concept of a fluid antenna system (FAS) over correlated alpha-mu fading channels, nicely fitting the THz communication. Furthermore, the considered system is expanded to the selection-combining-FAS (SC-FAS) and maximum-gain-combining-FAS (MGC-FAS) diversity variates at the receiver side. 
The proposed FAS and its diversity configuration techniques are aimed to combat the high path loss, blockages, and molecular absorption effect related to the THz band. Our contribution includes comprehensive outage probability (OP) performance analysis for the THz band given the non-diversity and diversity FAS receivers. Moreover, the derived outage probability formulas are verified via Monte Carlo simulations. Numerical results have confirmed the superior performance of the MGC-FAS scheme in terms of OP. Finally, this work justifies that a higher number of antenna ports dramatically improves the system performance, even in the presence of correlation. 
\end{abstract}
	
\IEEEpeerreviewmaketitle
	
\section{Introduction}
The research toward future communication systems and networks reasonably anticipates the movement of an operating frequency range from the millimeter wave (mmWave) to the terahertz (THz) band that ranges $0.1-10$ THz to meet the requirement of exponentially growing data demand. 
This transition provides endless possibilities for ultra-low latency, high-resolution sensing capabilities, and Tbps rates. THz communication enables remarkable applications such as connected robots, autonomous systems, extended reality, digital twins, and holographic teleportation \cite{7THz}.
However, THz communication requires new network architectures and device configurations to make all the remarkable features real. The authors in \cite{Antenna} analyzed the novel and potential antenna types for the THz band and characterized them as low-damage, wide-band, narrow-beam, and with high spectral resolution. The main challenge of novel THz antennas is the non-maturity of manufacturing materials and technology. 
%Moreover, the semiconductor equipment cannot easily transform electrical energy into electromagnetic at the THz spectrum. 

The authors in \cite{BruceLee} suggested that a software-controlled fluid antenna may replace multiple-input multiple-output (MIMO) antennas for the future generation networks and unleash remarkable properties of fluid antennas such as mechanical flexibility, theoretically infinite diversity, reduced electromagnetic radiation exposure, and exceptional immunity against multi-user interference.
The fluid antenna materials, fabrication, and potential applications were discussed in \cite{fluid}. The authors suggested that non-conventional methods, that are utilized to build Fluid-antenna-systems (FAS) like 3D printing, injecting, or spraying the conductive fluid material on substrates, increase the potential for reconfigurability. The authors in \cite{cst} presented a stretchable liquid-metal reconfigurable monopole antenna operating at 2.4 GHz that was designed on the CST program. The authors controlled the resonant frequency by stretching the antenna size and documented 1.8 GHz bandwidth reconfigurability. 

The FAS-enabled receivers were studied in \cite{WongFAS, WongNov2020} for a point-to-point network over correlated Rayleigh fading channels; the outage probability (OP) formulas for FAS-Rx was presented and the FAS performance was compared to the traditional maximum-ratio-combining diversity. The authors in \cite{Leila} applied a similar system model for an analog beamformed mmWave network and derived the OP formula in a single integral form. The authors in \cite{WongFAS,WongNov2020, Wong2022, Leila} reported a high performance of the FAS receivers even at a half wavelength size of an antenna length. It is expected that the FAS receivers can obtain more advantages for THz devices. Since the wavelength at THz is in the range of $0.03-3$ mm, it enables the installation of hundreds of antenna elements on a mobile device, whereas mmWave handheld devices could accommodate up to $32$ antenna elements.% One of the important advantages of THz spectrum is safety to environment and health of living beings \cite{}. According to \cite{EUreport}, previously employed FR1 frequency range is potentially carciogenic to humans. A new research by [4] reveals the safety of THz communication and therefore THz frequency range is used for medical scanning.

Channel modeling is one of the fundamental studies of 6G. The authors in \cite{Guan} presented a novel framework for modeling realistic THz channels with limited channel-sounding measurements. The proposed paradigm in \cite{Guan} consists of six phases such as limited channel sounding, calibration of ray tracing simulator, ray tracing simulations, stochastic channel model based on ray, cluster-based channel modeling. 
The channel measurements at $140$ GHz and $220$ GHz were conducted for indoor scenarios in \cite{Chen2021}, where the authors proposed a single-band close-in path loss model. Moreover, the authors in \cite{Chen2021WC} presented joint multi-path combining clustering and ray-tracing methods to study a signal propagation of THz waves in the indoor scenario and developed a hybrid channel model that applied both ray tracing and statistical methods to model the THz indoor channel. Similarly, the works in \cite{In1,In2,In3,In4,In5,In6} studied indoor THz channel modelling.
The authors in \cite{Nature} investigated popular fading channels with line-of-sight (LOS) components as Rician, Nakagami-$m$, and $\alpha-\mu$ distributions and compared them with empirical measurements at $142$ GHz frequency at different environments (i.e., airport, university, and shopping mall). It was proven that the $\alpha$-$\mu$ distribution well models the empirical data in the considered environments. 

In this work, we study a single-antenna transmitter (Tx) and FAS-enabled receiver with the correlation existing between antenna ports over THz channels. The contributions of this paper are outlined as follows:
\begin{itemize}
	\item The outage probability is evaluated for the FAS-enabled receivers at practical THz channels by considering a 3D propagation model and the molecular absorption loss.
	\item The cumulative distribution function (CDF) for the equally correlated $\alpha-\mu$ fading channels for the selection combiner output is presented in this work. To the best knowledge of authors, no prior work studied it before.		
	\item The diversity techniques for FAS-enabled receivers are introduced and their OP performance are investigated. 
\end{itemize}

\section{System Model}
\subsection{A Non-diversity FAS Receiver}
\begin{figure*}[!t]
	\centering
	\includegraphics[width=0.8\textwidth]{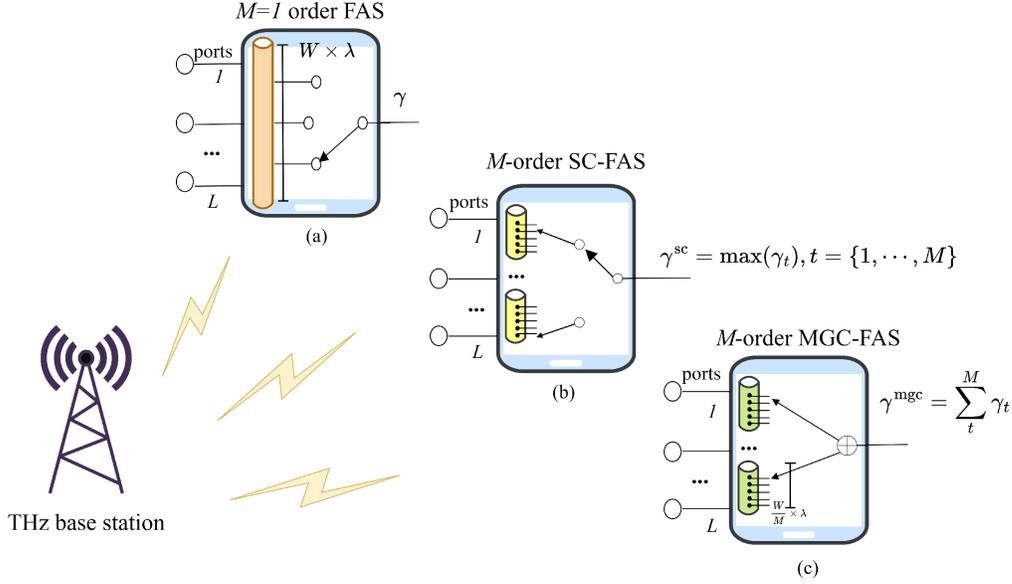}
	\caption{Schematic of a FAS-aided THz network with the SC-FAS and MGC-FAS diversity configurations.}
	\label{fig:system1}
\end{figure*}
%\begin{figure}[!t]
%	\centering
%	\includegraphics[width=0.9\linewidth]{System model}
%	\caption{\small Schematic of a FAS-aided THz network with the EGC-FAS and SC-FAS diversity configurations.
%		\vspace{-0.4cm}}
%	\label{fig:system1}
%\end{figure}

Assume a THz base station (BS) with a single directional transmit (Tx) antenna communicating to a single FAS receiver (Rx) with $L$ number of antenna ports. FAS-Rx is a tube filled with a liquid metal of the $W \times \lambda$ size, where $W>0$ is an antenna size coefficient and $\lambda$ is a wavelength of the operating frequency. Due to the mobility of the Rx device, there is a Doppler shift effect that causes a correlation between antenna ports. According to \cite{Wong2022}, the correlation coefficient is evaluated by a closed-form expression over all the ports as
\begin{align}
	r=\frac{2}{L (L-1)}\sum_{k=1}^{L-1} (L-k)J_0\left(\frac{2\pi  k W}{L-1}\right),
\end{align}
where $J_{0}$ is the zero-order Bessel function of the first kind. 
Hypothetically, we assume that FAS-Rx contains $L$ ports, mimicing antennas. A voltage controller enables switching to one of the ports, with the highest signal magnitude. For analytical tracktability, we do not account for the switching delay of a fluid material.

We consider an independent and identically distributed (i.i.d.)  $\alpha-\mu$ channel magnitude $|H_{k}|$ at $k$ port of FAS-Rx. The maximum selected channel is denoted as $|H_{\rm {max}}|=\displaystyle\max_{1<k<L} \{|H_{k}|\}$. The CDF of an $\alpha-\mu$ fading channel with envelop $|H_{k}|$, an arbitrary parameter $\alpha>0$, which is related to the non-linearity of the propagation medium, and $\alpha$ root mean value $\beta=(E(|H_{k}|^{\alpha}))^{\frac{1}{\alpha}}$ is given in \cite{Yacoub} as
\begin{align}
	F_{H_{k}}(x)=\frac{\Gamma(\mu,\frac{x^{\alpha}\mu}{\beta^{\alpha}})}{\Gamma(\mu)},
\end{align}
where $\mu>0$ is the inverse of normalized variance of $H_{k}$, $\mu=\frac{E^{2}((|H_{k}|^{\alpha}))}{Var(|H_{k}|^{\alpha})}$, with $E(\cdot)$ and $Var(\cdot)$ denoting the expectation and variance operators, and $\Gamma(x) = \int_{0}^{\infty}t^{x-1}e^{-t}{\rm d}t$. 
The correlated $\alpha-\mu$ channel amplitude is modelled as
\begin{align}\label{H}
	H_{k}=\left(\sum_{j=1}^{\mu}|h_{kj}|^{2}\right)^{\frac{1}{\alpha}},
\end{align} 
where $|h_{kj}|^{2}$ is parameterized below
\begin{multline}
	\label{h}
	h_{kj} = \left( \sqrt{1-r}X_{kj} + \sqrt{r}X_{0j}\right) \\ 
	+ i\left(\sqrt{1-r}Y_{kj}+\sqrt{r}Y_{0j} \right),
\end{multline}
where $k~=~\{1, \cdots, L\}$ and $j~=~\{1, \cdots, \mu\}$. $X_{0j}$, $X_{kj}$, $Y_{0j}$, and $Y_{kj}$ are the Gaussian random variables (RVs), with zero mean and $1/2$ variance. Furthermore, let us fix $U=\sqrt{r}(X_{0j}+iY_{0j})$ and re-write \eqref{h} as
\begin{align}
	h_{kj} &=\left( X_{kj}+i Y_{kj} \right) \sqrt{1-r}+U=\tilde{X}_{kj}+i \tilde{Y}_{kj},
\end{align}
where $\tilde{X}_{kj}=X_{kj}\sqrt{1-r}+\sqrt{r}X_{0j}$, $\tilde{X}_{kj}\sim N \left(\sqrt{r}Y_{0j}, \frac{1-r}{2}\right)$, $\tilde{Y}_{kj}=Y_{kj}\sqrt{1-r}+\sqrt{r}Y_{0j}$, $\tilde{Y}_{kj}\sim N \left(\sqrt{r}Y_{0j}, \frac{1-r}{2}\right)$, and $h_{kj} \sim C \left(0, \frac{1}{2}\right)$, $j~=~1, \cdots, \mu$.
Since the channel amplitude is the summation of squared magnitudes of Gaussian RVs, then, $H_{k}^{\alpha}$ is the non-central chi-square distribution represented by $H_{k}^{\alpha}\sim\chi_{2 \mu}\left(\sqrt{r\sum_{j=1}^{\mu}(x_{0j}^{2}+y_{0j}^{2}}), \frac{1-r}{2}\right)$, $j~=~\{1, \ldots, \mu\}$ and $X_{0j}~=~x_{0j}$, $Y_{0j}~=~y_{0j}$. The CDF of $\chi_{n}(a,b)$ is given in \cite[(2.1.124)]{Chi}, where $a^{2}$ is non-centrality parameter and $n$ is the degree of freedom.

%	\pmb{\text{OP}=\int _0^{\infty }\left(1-Q_m\left(\frac{\sqrt{2 r t }}{\sqrt{1-r}},\frac{\sqrt{2 \text{th}}}{\sqrt{\gamma (1-r)}}\right)\right){}^L\frac{t^{m-1}e^{\frac{-t}{2\sigma^2 }}}{\left(2 \sigma ^2 \right)^m\Gamma (m)}dt
	
\subsection{Propogation Model}
In this work, we have adopted a 3D signal propagation model by \cite{Shafie2020} that accounts for the molecular absorption loss. According to the authors in \cite{molecular}, THz frequencies suffer from the detrimental effect of molecular absorption loss that deteorates the received signal but also boosts the noise.

The received signal power is formulated as
\begin{align}
	\label{P}
	P=\frac{P_{t}G_{1}G_{2}c^{2}}{16 \pi^{2} f^{2}(d^{2}+(h_{b}-h_{u})^{2} \exp(K(f)(d^{2}+(h_{b}-h_{u})))},
\end{align}
where $P_{t}$ is the transmit power, $G_{1}$ is the BS's antenna gain, $G_{2}$ is the FAS-Rx gain, $c$ is the speed of light, $f$ is the operating frequency, $d$ is the distance between the BS and user equipment (UE), $h_{b}$ is the BS height, $h_u$ is the UE's height, $K(f)$ is the molecular absorption loss factor of the transmission medium depending on the operating frequency.
	
%\begin{}
\section{Outage Probability}
In this section, we evaluate the outage probability for the FAS-enabled THz networks. For this reason, we should quantify the CDF of the maximum signal-to-noise ratio (SNR) selection in equally correlated $\alpha-\mu$ distribution as
\begin{align}
	\label{gammafas}
	\gamma^{\rm fas} = \max(\gamma_1, \gamma_{2}, \ldots, \gamma_{L}).
\end{align}
The instantaneous SNR $\gamma_{k}$ is evaluated as
\begin{align}
	\gamma_{k}=\frac{P H_{k}}{N_{0}}, \quad k=1, 2, \ldots, L,
\end{align}
where $N_{0}$ is the additive white Gaussian noise (AWGN) per port.

The CDF of equally correlated $\alpha-\mu$ variates is generalized by making use of \eqref{H} as
{\allowdisplaybreaks
\begin{align}
	F_{H_{k}|U}(x|u) &= \Pr(H_{k}\leq x|u)\nonumber\\
	&=\Pr\left(\left(\sum_{j=1}^{\mu}|h_{kj}|^{2}\right)^{\frac{1}{\alpha}}<x\right) \nonumber\\
	&=\Pr\left(\sum_{j=1}^{\mu}|h_{kj}|^{2}< x^{\alpha}\right)\nonumber\\
	&=1-Q_{\mu}\left(\sqrt{\frac{2 r u}{1-r}}, \sqrt{\frac{2 x^{\alpha}}{(1-r)\bar{\beta}}}\right),
\end{align}}where $Q_{\mu}(\cdot, \cdot)$ is the Marcum-Q function and $\bar{\beta}=\frac{\beta^{\alpha}}{\mu}$. 
Now, we evaluate the CDF for the FAS output conditioned to the fixed port $U\sim \chi_{2 \mu}(0, \frac{1}{2})$, $U=\sum_{j=1}^{\mu}(X_{0j}^{2}+Y_{0j}^{2})$. 
The conditional CDF is represented as 
\begin{align}
	\label{CDF}
	&F_{\gamma_{\rm sc}|U}(x|u)=\Pr(\gamma_{1}\leq x^{\alpha}, \cdots, \gamma_{L}\leq x^{\alpha} |u) \nonumber\\
	&=\Pr\left(\sum_{j=1}^{\mu}|h_{1j}|^{2}\leq \frac{x^{\alpha}}{\bar{\gamma}}, \cdots, \sum_{j=1}^{\mu}|h_{Lj}|^{2}\leq\frac{x^{\alpha}}{\bar{\gamma}}|u\right)\nonumber\\
	&=\left(1-Q_{\mu}\left(\sqrt{\frac{2 r u}{1-r}}, \sqrt{\frac{2 x^{\alpha}}{(1-r)\bar{\beta}\bar{\gamma}}}\right)\right)^{L},
\end{align}
where $\bar{\gamma}=\frac{P E(H_{k})}{N_{0}}=\frac{P}{N_{0}}$ is the average SNR.
Now, by using the probability density function (PDF) of $U$ which is the central chi-square distribution with $2\mu$ degrees of freedom, $U \sim \chi_{2\mu} (0, \frac{1}{2})$. By averaging the conditional CDF in \eqref{CDF} over the distribution of 
$f_{U}(u)=\frac{u^{\mu-1}e^{-\frac{u}{2\sigma^{2}}}}{\Gamma(\mu)}$, we get the CDF of the FAS receiver for equally correlated $\alpha-\mu$ distribution as
\begin{align}
	\label{CDF1}
	F_{\rm sc}(x) &=\int\limits_{0}^{\infty}F_{\rm fas|U}(x|u)f_{U}(u){\rm d}u\nonumber\\
	&=\!\int\limits_{0}^{\infty}\!\left(1\!-\! Q_{\mu}\left(\sqrt{\frac{2 r u}{1-r}}, \sqrt{\frac{2 x^{\alpha}}{(1-r)\bar{\beta}\bar{\gamma}}}\right)\right)^{L} \nonumber \\
	&~~~\times \frac{u^{\mu-1}e^{\frac{-u}{2\sigma^{2}}}}{(2\sigma^{2})^{\mu}\Gamma(\mu)}{\rm d}u.
\end{align}
Finally, the outage probability for FAS-enabled Rx is obtained by using \eqref{CDF1} $OP(x)=F_{\rm sc}(x)$.

\subsection{Diversity FAS Receivers}
In this section, we propose the $M$-order SC-FAS and MGC-FAS diversity techniques and evaluate their outage probability formulas. Fig. 1 illustrates the top view of the non-diversity FAS (a), SC-FAS (b), and MGC-FAS (c). The traditional MGC-combiner technique consider a weighted summation of SNR-branches. In the proposed MGC-FAS technique, we assume that SNR branches are added without being divided into the diversity order. The FAS outputs deterministic signal and MGC-FAS sums the SNR from diversity branches. We assume that a FAS tube with $W\times \lambda$ size is divided into $M$ sub-FAS branches with size $\frac{W\times\lambda}{M}$. $M$-order SC-FAS and MGC-FAS device contain $\frac{L}{M}$ antenna ports per FAS branch. 
The outage probability for the Rx-enabled SC-FAS is defined by using \eqref{CDF1} and by adjusting the antenna ports to $\frac{L}{M}$ as
{\allowdisplaybreaks
\begin{align}
	&OP^{\rm sc}(x)=\left(F_{\rm sc}(x)\right)^{M}\nonumber\\
	&=\!\left(\!\int\limits_{0}^{\infty}\hspace{-0.1cm}\left(\hspace{-0.1cm}1\!-\! Q_{\mu}\hspace{-0.1cm}\left(\hspace{-0.1cm}\sqrt{\frac{2 r u}{1-r}}\!,\!\sqrt{\frac{2 x^{\alpha}}{\bar{\gamma}(1-r)}}\!\right)\hspace{-0.1cm}\right)^{\frac{L}{M}}\hspace{-0.2cm}\frac{u^{\mu-1}e^{\frac{-u}{2\sigma^{2}}}}{(2\sigma^{2})^{\mu}\Gamma(\mu)}{\rm d}u\!\right)^{M}\!.\!
\end{align}}

Similarly to the traditional maximum-ratio combining technique, the proposed FAS-MGC diversity technique sums the output SNRs from every FAS tubes as $\gamma_{\rm fas}^{\rm mgc}=\sum_{t=1}^{M}\gamma^{\rm fas}_{t}$. The evaluation of the outage probability for the correlated MGC-FAS receivers is mathematically intractable, therefore, we have obtained the outage probability performance by doing Matlab simulations.
\begin{align}
	&OP^{\rm mgc}(x)=\Pr(\gamma_{1}^{\rm fas}+\gamma_{2}^{\rm fas}\cdots +\gamma_{M}^{\rm fas}\leq x^{\alpha}),
\end{align}
where $\gamma_{t}^{\rm fas}$ is defined in \eqref{gammafas}.
%Next, the outage probability for the MGC-FAS   is evaluated by 
%\begin{align}
%	&OP^{\rm egc}(x)=\nonumber\\
%	&\int\limits_{0}^{\infty}\left(1-Q_{\mu}\left(\sqrt{\frac{2 r u}{1-r}}, \sqrt{\frac{2 \gamma_{th}^{\alpha}}{\sum_{t=1}^{L/M}\bar{\gamma}_{t}(1-r)}}\right)\right)^{M}\nonumber\\
%	&\times\frac{u^{\mu-1}e^{\frac{-u}{2\sigma^{2}}}}{(2\sigma^{2})^{\mu}\Gamma(\mu)}{\rm d}u.
%\end{align}
\begin{figure}[!t]
	\centering
	\includegraphics[width=\linewidth]{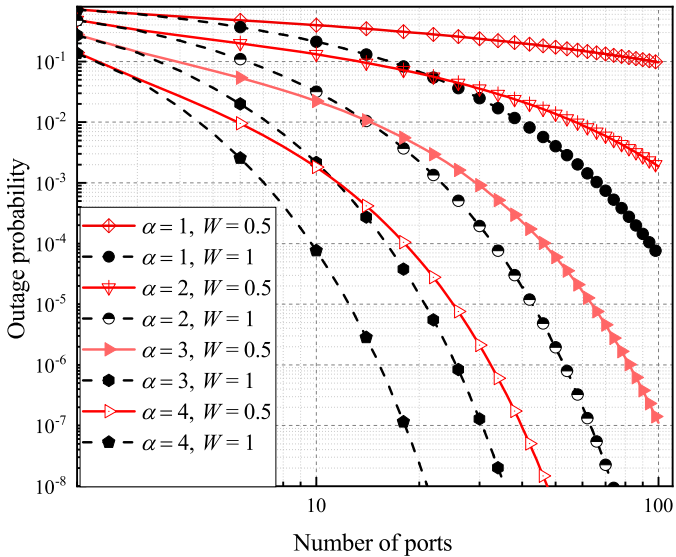}
	\caption{Outage probability versus the number of ports for the FAS-aided THz network for $\alpha~=~\{1, 2, 3, 4\}$ and $W~=~\{0.5, 1\}$ Rx-FAS.}
	\label{fig:fig1}
\end{figure}

\section{Numerical Simulation}
This section presents simulation results obtained from analysis of the outage probability for the FAS-enabled receivers at the THz frequency range. We simulate the THz channel with correlated FAS ports given the following system parameters: Tx power $P_{t}~ =~ 20$ dBm, Tx antenna gain $G_1 ~=~
17$ dBi, Rx antenna gain $G_2 ~=~ 14$ dBi, Tx height $h_1 ~=~ 4$ m, Rx height $h_2 ~=~ 1$ m, distance between Tx-Rx is $d ~=~ 10$ m, operating frequency $f ~=~ 1$ THz, bandwidth is $B ~=~ 10$ GHz, $R~=~7$ Gbits/s, $K(f)~=~0.192$ $\text{m}^{-1}$ as in \cite{Shafie2020}, and the range of FAS ports are $L ~=~ \{2, 100\}$.
\begin{figure}[!t]
	\centering
	\includegraphics[width=\linewidth]{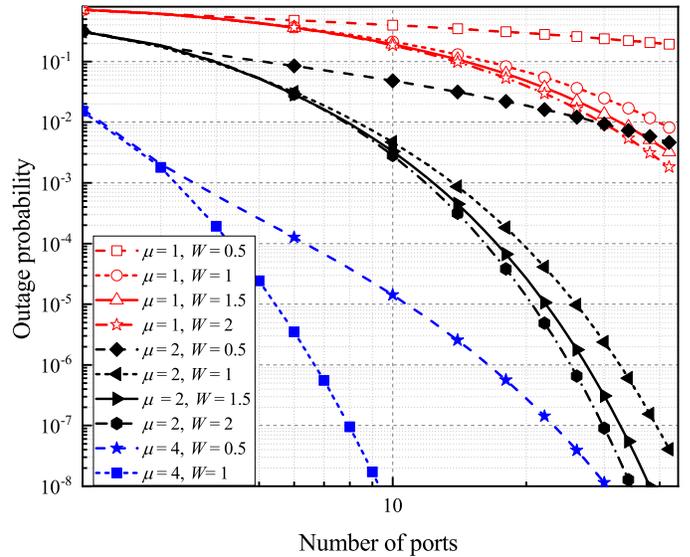}
	\caption{Outage probability versus the number of ports for the FAS-aided THz network for $\mu~=~\{1, 2, 4\}$ and $W~=~\{0.5, 1, 1.5, 2\}$ Rx-FAS.}
	\label{fig:fig2}
\end{figure}

In Fig. 2, we analyze how varying $\alpha~=~\{1, 2, 3, 4\}$ power parameter and antenna size coefficient $W~=~ \{0.5, 1\}$ affect the outage probability at $\mu~=~1$.
Let us consider plots for $L~=~50$ antenna ports at $W~=~1$ and record the following results $OP(\alpha=1)~=~0.237$, $OP(\alpha=2)~=	0.035$, $OP(\alpha=3)~=~9.085\times 10^-{-4}$,  $OP(\alpha=4)~=~2.115\times 10^{-6}$. A higher $\alpha$ value results in a stronger outage probability performance. In addition, we notice that the outage probability performance stays almost the same when $L<4$ for $W~=~0.5$ and $W~=~1$ cases. However, when the number of ports increases, there is a noticeable outage probability difference due to the increasing correlation coefficient between ports. 

In Fig. 3, we study how channels parameter $\mu~=~ \{1, 2, 4\}$, which identifies the number of multipath clusters, along with the antenna size coefficients $W ~=~ \{0.5, 1, 1.5, 2\}$ at the fixed $\alpha=1$ affect the outage probability performance. Given the same antenna size coefficient, $W ~=~ 0.5$ and different $\mu$-parameter values produce a substantial difference in the OP performance. For instance, for
$L = 10$ ports, the outage probability of $OP(\mu=1)~=~0.42$, $OP(\mu=2) ~= ~0.047$ , $OP(\mu=4)~=~1.5 \times 10^{-5}$ are recorded at $\mu = \{1, 2, 4\}$. It means that a higher number of multipath clusters significantly increases the outage probability. At the same time, the smallest antenna size
coefficient $W~=~ 0.5$ results in a lower outage probability performance in comparison to other
counterparts due to a higher effect of correlation between ports. For instance, at $L~=~20$ and $\mu~=~2$ outage probability is recorded as
$OP(W=0.5)~=~0.0187$, $OP(W=1)~=~8.574\times 10^{-5}$, $OP(W=1.5)~=~2.653\times 10^{-5}$, $OP(W=2)~=~1,357\times 10^{-5}$. Hence, there is $218$ times performance improvement from $W=0.5$ plot to $W~=~1$ case. Moreover, this figure demonstrates that an increasing number of antenna ports dramatically enhances the outage probability for all scenarios.
Specifically let us consider plots for $\mu~=~2$ and $W~=~0.5$ the  OP is obtained as $OP(L=10)~=~0.048$, $OP(L=20)~=~0.018$, $OP(L=30)~=~0.009$, $OP(L=40)~=~0.005$, $OP(L=50)~=~0.003$, $OP(L=60)~=~0.002$, and $OP(L=100)~=~3.703 \times 10^{-4}$. So, from $L~=~10$ to $L~=~100$ case there is a $130$ times outage probability performance enhancement.
\begin{figure}[!t]
	\centering
	\includegraphics[width=0.9\linewidth]{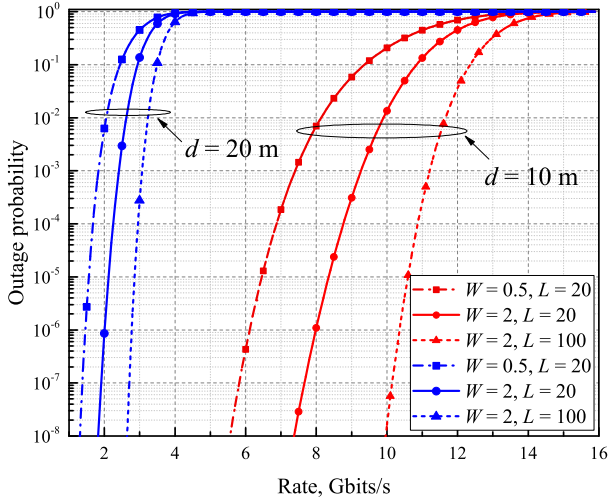}
	\caption{Outage probability  versus rate for the FAS-aided THz network for $\mu~=~2$, $W~=~\{0.5, 2\}$, and $d~=~10, 20$ m Rx-FAS.}
	\label{fig:fig3}
\end{figure}

In Fig. 4, we analyze how the distance between the base station and receiver, antenna size, and the number of antenna ports reflect on the outage probability. For this reason, we consider the following environment $\alpha~=~2$, $\mu~=~2$, and the Rx-FAS with $L~=~\{20, 100\}$ and $W~=~\{0.5, 2\}$. If we consider $L~=~20$ and $d~=~10$ m scenario for the  $W~=~\{0.5, 2\}$ FAS, then we observe around $2$ Gbits/s rate difference from a $W~=~0.5$ to $W~=~2$ case. Similarly, we observe $6$ Gbits/s rate improvement from $L~=~20$ to $L~=~100$ antenna ports per receiver FAS. In other words, a higher number of antenna ports has a higher impact on the network performance. Similar, observations are made for the $d~=~20$ m plots. 
Furthermore, the distance is a crucial factor in the THz range, even $10$ m difference creates $4$ times decrease in achievable rate to meet $OP ~=~ 10^{-4}$ at $W~=~ 2$ and $L ~=~ 20$ scenario. For instance, at $L~=~ 20$ m and $W ~=~ 2$ to obtain performance of $OP ~=~ 10^{-4}$, we obtain following rates at $10$m and $20$m distances as $R(10 \text{m})~=~8.8$ Gbits/s and $R(20 \text{m})~=~2.37$ Gbits/s. Furthermore, this figure demonstrates that one could combat a high path loss at the THz range by using FAS with a higher number of ports. 

In Fig. 5, we compare the $M~=~\{2, 4\}$-order MGC-FAS and SC-FAS with the non-diversity FAS receiver for $L~=~100$ ports. It means that for $M~=~2$-order FAS there are two FAS tubes with $L~=~50$ ports each. For simulation purposes we use $W~=~2$, $\mu~=~2$, $\alpha~=~2$, $R~=~9$ Gbits/s, and $d~=~13$ m. For clarity purposes, we consider two sub-plots for $M~=~2$ (a) and $M~=~4$ diversity orders of SC-FAS and MGC-FAS.
 %We consider practical THz channel environment parameter measured from 'shopping mall' environment as in \cite{Nature}, $d~=~10.04$ m, $\alpha~=~3.28199$, and $\mu~=~1.71725$. We have sub-ploted
 There are several important observations made from this figure. The MGC-FAS diversity scheme demonstrates the highest performance among all plots. As the number of diversity branches increases in the MGC-FAS the performance also significantly improves. For instance, $M~=~4$-order MGC-FAS with $L=8$ results in $OP(M=4)~=~0.3368$, whereas, $M~=~2$-order MGC-FAS returned $OP^{\rm mgc}(M=2)~=~0.8745$. Oppositely, the SC-FAS scheme demonstrates a lower performance when the number of diversity orders rises from $M~=~2$ to $M~=~4$. For example, $OP^{\rm sc}(M=2, L=32)~=~0.68$ and $OP^{\rm sc}(M=4, L=32)~=~0.072667$. Next, the $L$-port FAS slightly outperforms the $M~=~2$-order SC-FAS scheme.
 
 % outperforms the FAS one, when the number of ports is $L<31$ for $M=6$-order SC-FAS, $L<47$ for $M=4$-order SC-FAS. Moreover, $M=2$-order SC-FAS outperformance the non-diversity FAS for all ports. It appears that $M=2$-order SC-FAS demonstrates the optimum performance among all plots.
\begin{figure}[!t]
	\centering
	\includegraphics[width=1\linewidth]{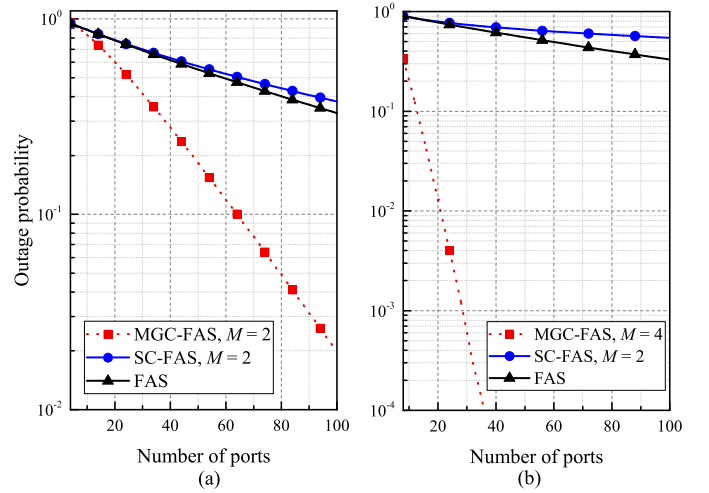}
	\caption{Outage probability  versus the number of ports for the non-diversity FAS, MGC-FAS, and SC-FAS.}
	\label{fig:fig4}
\end{figure}

\section{Conclusion}
One of the candidate technologies for $6$G is the THz communication ranging in ($0.1-10$ THz) spectrum.
In this work, we aimed to model a THz channel by considering 3D propagation model of the signal and molecular absorption effect of THz band. We suggested using the FAS-enabled receivers to improve the end-user performance. The proposed model is analytically tractable and provides important insights in the analysis of THz networks. We analyzed both non-diversity and diversity FAS receivers in terms of the outage probability formulas. Simulation results demonstrated that a non-diversity FAS outperforms the SC-FAS, however, it demonstrates a significantly lower performance rather than the MGC-FAS scheme. Moreover, simulation results reveal that the higher values of $\alpha$ and $\mu$ channel parameters, more antennas and antenna ports improve the system performance. 

\balance

\bibliographystyle{IEEEtran}
\end{document}